\documentclass[aps, pra, twocolumn, unsortedaddress, noeprint]{revtex4-1}
\usepackage{graphicx}
\usepackage{hyperref}
\usepackage{epstopdf}
\usepackage{makeidx}
\usepackage{amsmath}
\usepackage{color}
\usepackage{bm}
\usepackage[english]{babel}
\usepackage{dcolumn}
\usepackage{latexsym} 
\graphicspath{{figure/}}
\makeindex
\begin{document}
\title{Programming multi-level quantum gates in disordered  computing reservoirs via machine learning and TensorFlow}
\author{Giulia Marcucci}
\affiliation{Department of Physics, University Sapienza, Piazzale Aldo Moro 2, 00185 Rome (IT)}
\affiliation{Institute for Complex Systems, National Research Council (ISC-CNR), Via dei Taurini 19, 00185 Rome (IT)}
\author{Davide Pierangeli}
\affiliation{Department of Physics, University Sapienza, Piazzale Aldo Moro 2, 00185 Rome (IT)}
\affiliation{Institute for Complex Systems, National Research Council (ISC-CNR), Via dei Taurini 19, 00185 Rome (IT)}
\author{Pepijn Pinkse}
\affiliation{Complex Photonic Systems (COPS), MESA+ Institute for Nanotechnology, University of Twente, P.O. Box 217, 7500 AE Enschede, The Netherlands}
\author{Mehul Malik}
\affiliation{Institute of Photonics and Quantum Sciences (IPAQS), Heriot-Watt University, Edinburgh, EH144AS UK}
\author{Claudio Conti}
\affiliation{Department of Physics, University Sapienza, Piazzale Aldo Moro 2, 00185 Rome (IT)}
\affiliation{Institute for Complex Systems, National Research Council (ISC-CNR), Via dei Taurini 19, 00185 Rome (IT)}
 \email{claudio.conti@uniroma1.it}
 \date{\today}

 \begin{abstract}
   Novel machine learning computational tools open new perspectives for quantum information systems. Here we adopt the open-source programming library TensorFlow to design multi-level quantum gates including a computing reservoir represented by a random unitary matrix. In optics, the reservoir is a disordered medium or a multi-modal fiber. We show that trainable operators at the input and the readout enable one to realize multi-level gates. We study various qudit gates, including the scaling properties of the algorithms with the size of the reservoir. Despite an initial low slop learning stage, TensorFlow turns out to be an extremely versatile resource for designing gates with complex media, including different models that use spatial light modulators with quantized modulation levels.
 \end{abstract}


\maketitle
\section{Introduction}
The development of multi-level quantum information processing systems has steadily grown over the past few years, with experimental realizations of multi-level, or qudit logic gates for several widely used photonic degrees of freedom, such as orbital-angular-momentum and path encoding \cite{Huisman:15, Babazadeh2017, Malik2016, Taballione:19}.
However, efforts are still needed for increasing the complexity of such systems while still being practical, with the ultimate goal of realizing complex large-scale computing devices that operate in a technologically efficient manner.

A key challenge is the development of design techniques that are scalable and versatile. Recent work outlined the relevance of a large class of devices, commonly denoted as ``complex'' or ``multi-mode''
\cite{Matthes2018,Zhao2018}.
In these systems, many modes or channels are mixed and controlled at input and readout to realize a target input-output operation.
This follows the first experimental demonstrations
of assisted light transmission through random media \cite{Vellekoop2007,Vellekoop2015,Yang2015,Pierangeli2018}, which demonstrated many applications including arbitrary linear gates \cite{Matthes2018}, mode conversion, and sorting \cite{Fu2018, Carpenter2018}.

The use of complex mode-mixing devices is surprisingly connected to leading paradigms in modern machine learning (ML), as the ``reservoir computing'' (RC) \cite{Verstraeten2007}, and the ``extreme learning machine'' (ELM) \cite{Huang2006,Verstraeten2007}. 
In standard ML, one trains the parameters ({\it weights}) of an artificial neural network (ANN) to fit a given function, which links input and output. In RC, due to the increasing computational effort to train a large number of weights, one internal part of the network is left untrained (``the reservoir'') and the weights are optimized only at input and readout.

ML concepts, such as photonic neuromorphic and reservoir computing \cite{Duport2012, Van2017}, are finding many applications in telecommunications \cite{Borhani18,Lohani18}, multiple scattering \cite{Sun2018}, image classification \cite{Lin2018}, metasurfaces \cite{Fratalocchi2018, Engheta2019},  biophotonics \cite{Pierangeli2018}, Ising machines \cite{Pierangeli2019},  integrated and fiber optics \cite{Wu2014, Englund2017}, and topological photonics \cite{Pilozzi2018}. Various authors have reported the use of ML for augmenting and assisting quantum experiments \cite{Krenn2016, Marquardt2018,Sciarrino2018, Shalaev2019, Gigan2019}. The field of machine learning is in turn influenced by quantum physics, for example in the orthogonal units \cite{Jing2017, Dangovski2017}.

Here we adopt RC-ML to design complex multi-level gates \cite{Babazadeh2017,Malik2016,Stroud2000,Di2011}, which form a building block for high-dimensional quantum information processing systems. While low-dimensional examples of such gates have been implemented using bulk and integrated optics, efficiently scaling them up to high dimensions remains a challenge. 

In quantum key distribution (QKD), one uses at least two orthogonal bases to encode information. High-dimensional QKD offers an increased information capacity as well as an increased robustness to noise over qubit-based protocols \cite{Mirhosseini2015,Ecker2019}. Such protocols may be realized by using the photonic spatial degrees of freedom as the encoding (computational) basis, and suitable unitary operators to switch between bases mutually unbiased with respect to the computational basis. However, the security of the QKD protocol may be compromised by the fidelity of such basis transformations, leading to errors in the key rate. An additional consideration is the experimental complexity of such transformations, which can scale rather poorly using established techniques based on bulk optical systems. By using a random medium and I/O readout operators, one can realize such high-dimensional operations in a controllable and scalable manner, relying only on the existing complexity of the disordered medium and a control operation at the input. Here, we explore methodologies to train a disordered medium to function as a multi-level logic gate by using different implementations of ML concepts.

\begin{figure*}
  \centering
  \includegraphics[width=2.0\columnwidth]{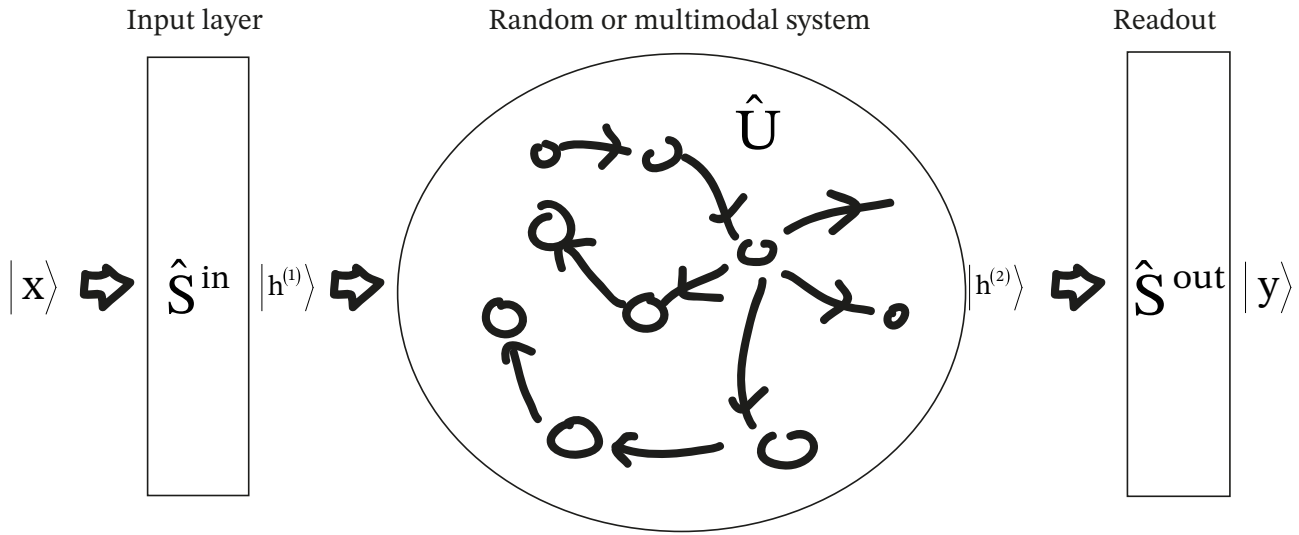}
  \caption{A general optical gate based on a complex
    random medium; the input state ${\bf x}$ is processed
    to the input layer with operator ${\hat{S}^{\text{in}}}$,
    the system is modeled by the unitary operator ${\hat U}$,
    and the output is further elaborated by ${\hat{S}^{\text{out}}}$.
  \label{fig:scheme}}
\end{figure*}

Figure \ref{fig:scheme} shows the schematic of a device including the
complex medium, represented by the unitary operator $\hat{U}$,
and two trainable input $\hat{S}^{\text{in}}$ and readout $\hat{S}^{\text{out}}$ operators. 
$|h^{(1,2)}\rangle$ are hidden states. The use of an optical gate in this manner is related to the use of a disordered medium as a physically unclonable function (PUF) \cite{Goorden2014, Tentrup2018,LeonettiKarbasiMafiEtAl2016}.

In our general framework, we have a random system modeled by a unitary random matrix. We want to use the random medium to perform a computation in a Hilbert space containing many qudits.
The random medium is not necessarily a disordered system (for example, a dielectric assembly of scattering particles), but may also be a multimode fiber, or an array of waveguides.
The input/output relation is represented by a linear unitary matrix operator
$U_M$ and only forward modes are considered. The $U_M$ matrix has dimensions $M\times M$, with $M$ the dimension of the {\it embedding space}.

The ``reduced'' state vector at input has dimensions $N\times 1$, with $N\leq M$. This models the case in which we use a subset of all the available modes.
The input to the reservoir is a ``rigged'' state vector ${\bf x}$ with dimension $M$, where the missing complementing $C$ components are replaced by $C=M-N$ {\it ancillas}.
Our goal is to use the random medium to perform a given operation denoted by a gate unitary matrix
\begin{equation}
T_M=S^{\text{out}}_M\cdot U_M \cdot S_M^{\text{in}}.
\label{generaleq}
\end{equation}
$S_M^{\text{in}}$ and $S_M^{\text{out}}$ are two ``training'' operators that are applied at input and output
(see Fig.~\ref{fig:scheme}) and whose elements can be adjusted.
We first consider the presence of the input operator $S_M^{\text{in}}=S_{M}$, and
$S_M^{\text{out}}={\bf 1}_M$, which can be implemented by spatial-light modulators
(we denote as ${\bf 1}_M$ the identity matrix with dimension $M$).

We identify two cases: either (i) we know the matrix $U_M$, or (ii) we have to infer $U_M$ from the input/output relation. We show in the following the way these two problems can be solved by ANNs, where we denote
the two families as {\it non-inferencing} and {\it inferencing} gates.
\newline \noindent
\section{Non-inferencing gates}
We consider a target gate with complex-valued input state with dimension $N$, and components $x_1,x_2,...,x_N$.
We embed the input vector in a rigged Hilbert space with dimension $M\geq N$,
so that the overall input vector is ${\bf x}=\{x_1,x_2,...,x_{N},x_{N+1},...,x_{M}\}$. We have a linear propagation through a medium with unitary complex transfer matrix $U_M$.
The overall transmission matrix is $T_M=U_M\cdot S_M$, such that the output vector is ${\bf y}=T_M\cdot {\bf x}=U_M\cdot\,S_M\cdot {\bf x}$. The observed output vector is written as $P\cdot {\bf y}$,
where $P$ is a $N-$projector operator with dimensions $N\times M$ such that $P=[ \bm{1}_N | {\bf 0} ]$, with $\bm{1}_N$  the identity matrix with size $N\times N$, and ${\bf 0}$ a null matrix with dimension $N\times C$. The goal is finding the matrix $S_M$ such that
\begin{equation}
  P\cdot U_M\cdot S_M=[X_N \,\vline\,  {\bf 0}]
  \label{eq1}
  \end{equation}
 where $X_N$ is the $N\times N$ target
  gate and ${\bf 0}$ is the null complement $N\times C$ at dimension $M$.
  Eq.~(\ref{eq1}) is a matrix equation, which guarantees that the
  overall system behaves as a $X_N$ gate on the reduced input.

  Solving the matrix Eq.~(\ref{eq1}) may be demanding and nontrivial
  when the number of dimensions grows.
  In the following, we discuss the use of ML  techniques.
  
The transmission matrix $T_M$ in the rigged space from ${\bf x}$ to   ${\bf y}$ can be written as blocks
  \begin{equation}
    \label{eq:2}
    T_M=
    \begin{bmatrix}
      X_N & \vline & {\bf 0} \\
      \hline
      {\bf 0} & \vline & O_{C}
      \end{bmatrix}
    \end{equation}
    where $O_C$ is a unitary matrix with dimensions  $C\times C$ to be determined. If $U_M$ and $S_M$ are unitary, the resulting transmission matrix $T_M$ is also unitary.
    However, if one uses Eq.~(\ref{eq1}), the problem may also have a nonunitary solution (``projected case'') as some channels are dropped at the output.
    In other words, solving Eq.~(\ref{eq:2}) is not equivalent to solving
    Eq.~(\ref{eq1}), and we adopt two different methodologies:
   one can look for unitary or nonunitary solutions by ANN.

\begin{figure*}
  \centering
  \includegraphics[width=2.0\columnwidth]{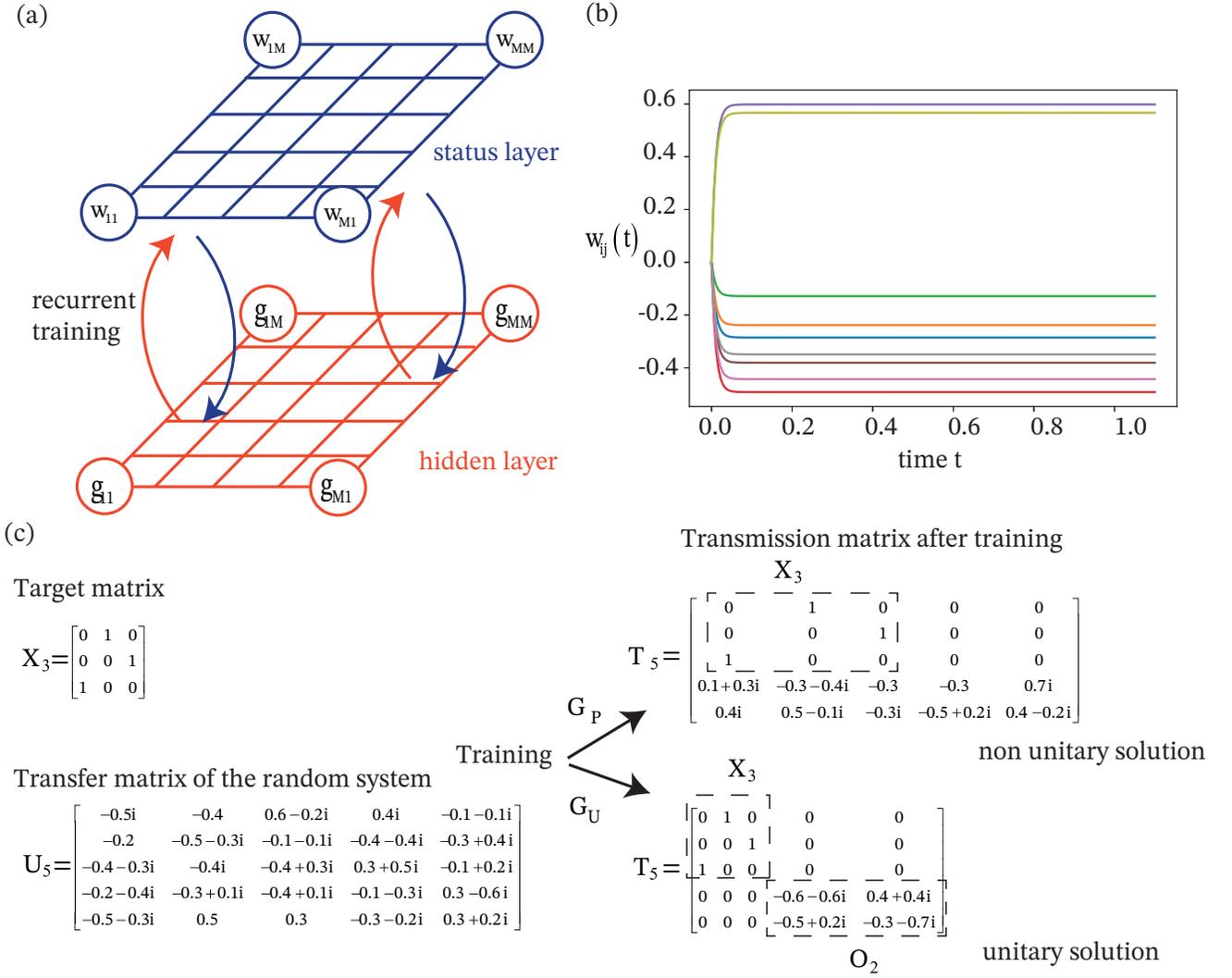}
  \caption{(a) Recurrent neural network for the matrix equation \eqref{eq:6}.
    The status nodes are denoted by the elements of the matrix $W$, and the
    hidden state of the system is in the nodes of the matrix
    $F$;  (b) training dynamics for the case $N=M=3$ with $X_T$ corresponding
to a single-qutrit X-gate ($\mu=100$);
  (c) resulting transfer function for the case $N=3$ and $M=5$ in the
  unitary and non-unitary case. In the latter case, the excess channels are ignored during the training.  The resulting transmission channels $T_M$ are displayed, $O_2$ is the unitary complements for $C=M-N=2$ in the unitary case.}
  \label{fig:rc}
\end{figure*}

By following previous work developed for real-valued matrices \cite{Wang1993}, we map the complex-valued matrix equation \eqref{eq1} into a recurrent neural network (RNN). In the ``non-inferencing'' case, the
matrix $U_M$ is known, and the solution is found by the RNN in Fig.~\ref{fig:rc}.
The RNN solves an unconstrained optimization problem, by finding the minimum of the sum of the elements $e_{ij}>0$ of an error matrix $E$. The error depends on a ``state matrix'' $W_M$, and one trains the elements $w_{ij}$  of $W_M$ to find the minimum
\begin{equation}
  \label{eq:5}
\underset{W_M}{\text{min}}  E[G(W_M)]=\underset{W_M}{\text{min}}\sum_{i,j} e_{ij}[G(W_M)]\text{.}
\end{equation}
In the adopted approach, the sum of the elements $e_{ij}$ is minimal when the {\it hidden layer} elements $g_{ij}$ of the matrix $G(W)$ are zero. $E$ and $G$
have to be suitably chosen to solve the considered problem.
We found two possible $G$ matrices: (i) the ``projected'' 
\begin{equation}
  G_P=P \cdot U_M \cdot W_M-X_{N0},
  \end{equation}
with $X_{N0}=[X_N \,\vline\,  {\bf 0}]$ as in Eq.~(\ref{eq1}) and, (ii) the ``unitary'' [see Eq.~\eqref{eq:2}]
  \begin{equation}
    G_U=U_M\cdot W_M-T_M.
  \end{equation}
These two cases are discussed below.

  To find the unknown training matrix $S_M$, one starts from an initial guess matrix $W_M(0)$. The guess is then recurrently
  updated, as in Fig.~\ref{fig:rc}, until a stationary state $W_M(\infty)$ is reached.  Once this optimization converges, the solution is given by $S_M=W_M(\infty)$. The update equation is determined by a proper choice
  of the error matrix $E$ as follows.

As the matrices are complex valued, $e_{ij}$ is a function
of $g_{ij}$ and $g_{ij}^*$. We set $e_{ij}=e_{ij}(|g_{ij}|^2)$.
The corresponding dynamic RNN equation, which
for large time gives the solution to the optimization problem, is
\begin{equation}
  \label{eq:6}
  \frac{d W_M}{dt}=-\mu U_M^{\dagger}\cdot F[G(W_M)]
\end{equation}
where $\mu$ is the ``learning rate'', an optimization coefficient
(hyperparameter), which is set to speed-up the convergence. 
The elements $f_{ij}$ of the matrix $F$ are $f_{ij}=\frac{d e_{ij}}{d g_{ij}^*}$.
Letting $e_{ij}=|g_{ij}|^2$, one has $f_{ij}=g_{ij}$.

Eq.~\eqref{eq:6} implies that the RNN is composed of two bidirectionally connected layers of neurons, the output layer
with state matrix $W$, and the hidden layer with state matrix $G$.
The training corresponds to sequential updates of $F$ and $W$ when solving
the ordinary equations [\eqref{eq:6}].
As shown in  \cite{Wang1993}, this RNN is asymptotically
stable and its steady state matrix represents the solution
(an example of training dynamics is in Fig.~\ref{fig:rc}b).

We code the RNN by TensorFlow\textsuperscript{TM} and use the
ordinary differential equations (ODEs) integrator {\it odeint}. In the case $N=M$, as $X_N=X_M$ is a unitary operator, the solution of the recurrent network furnishes a unitary $S_M$ matrix, which solves the problem. For $M>N$ the RNN furnishes a unitary solution $S_M$,
and a unitary transfer function $T_M$, only if  we embed the
target gate $X_N$ in a unitary operator as in $\eqref{eq:2}$ with $O_C$
a randomly generated unitary matrix.

\subsection{Single non-inferencing qutrit gate X }
For the training of a gate $X_3$ defined by \cite{Babazadeh2017, Krenn2019}
\begin{equation}
  X_3=\sum_{l=0}^{d-1}|l\oplus 1\rangle \langle l |=\begin{bmatrix}
  0 & 1 & 0 \\ 0 & 0 & 1 \\ 1 & 0 & 0
\end{bmatrix}\label{eq:1}
  \end{equation}
  The gate $X_3$ is obtained by an embedding dimension $M=5$ and
  unitary transfer function $U_5$ as in Fig.~\ref{fig:rc}.
  
For $G=G_P$, the number of ODEs for the training of the network is minimal ($N=3$). However, the solution is 
not unitary, as some channels are dropped out by the $N-$projector. The overall  $M\times M$ transmission matrix $T_M$, after the training, is such that $T_M^\dagger \cdot T_M\neq I$ because
the solution $S_M$ is not unitary. 
However, the system always reaches a stationary case.

A unitary solution is found by letting $G=G_U$ and involving the maximum number of ODEs in \eqref{eq:6} with a unitary embedding of $X_N$ as in $\eqref{eq:2}$, i.e., adopting a further - randomly generated - unitary matrix $O_C$.
The key point is that the system {\it finds a solution for any random unitary rigging of the matrix $X_N$}, that is, for any randomly assigned matrix $O_C$. This implies that we can train all these systems to realize different multi-level gates.

\noindent
\section{Inferencing gates}
In the case that we do not know the transfer matrix of the system, we can still train the overall transmission matrix by using a neural network and infer $U_M$.
Here we use an ANN to determine the training operators without measuring the transfer matrix. Figure \ref{fig:inference1} shows the scheme of the ANN, where
the unitary matrix $U_M$ is represented by its elements $u_{ij}$,
and the $w_{ij}$ are the adjustable weights.
After training, the resulting ${w}_{ij}$ are the elements of the solution matrix $S_M$.
For the sake of simplicity, we consider $S^{\text{out}}={\bf 1}_M$, as above.
For a target $X_N$ we build the $T_M$ as in (\ref{eq:2}) by randomly
generating the unitary complement $O_C$.
As $T_M$ and $U_M$ are unitary, the
resulting $S_M$ is also unitary.
One can use a non unitary $T_M$ by choosing, for example, $O_C={\bf 0}$. Correspondingly - after the training - $S_M$ is not unitary.

We randomly generate a set of input states ${\bf x}_i$, with $i=1,...,n_{train}$. Each input state is ``labelled'' with the target output
${\bf y}_i=T_M \cdot {\bf x}_i$. We remark that ${\bf x}_i$ and ${\bf y}_i$ are vector with size $M$.
A further set of $n_{valid}$ validation rigged vectors is used to validate
the training.

For any input ${\bf x}_i$ in the training set, we adjust the weights
to minimize the error function
\begin{equation}
  e_i=\frac{1}{N}\sum_{N} |{\bf y}_i-U_M\cdot W_M\cdot {\bf x}_i|^2
\label{eq:cost}
\end{equation}
with ${\bf y}_i=T_M\cdot {\bf x}_i$.
After this training, we test the accuracy on the validation set.
Each cycle of training and validation is denoted as ``epoch''.
 
Figure \ref{fig:inference1} shows the ANN for $N=3$, and $M=5$.
In our model, we build a matrix $W_M$ of unknown weights.
As we deal with complex quantities, $W_M$ is written
as $W_M=W_M'+\imath W_M''$ with $W_M'$ and  $W_M''$ real-valued matrices,
whose elements form the weights of the ANN.
Using random matrices as initial states,
we end the iteration when the validation cost is below a threshold $\varepsilon_{valid}$. 
\subsection{Single-qutrit inference X-gate}
\begin{figure*}
  \centering
 \includegraphics[width=\textwidth]{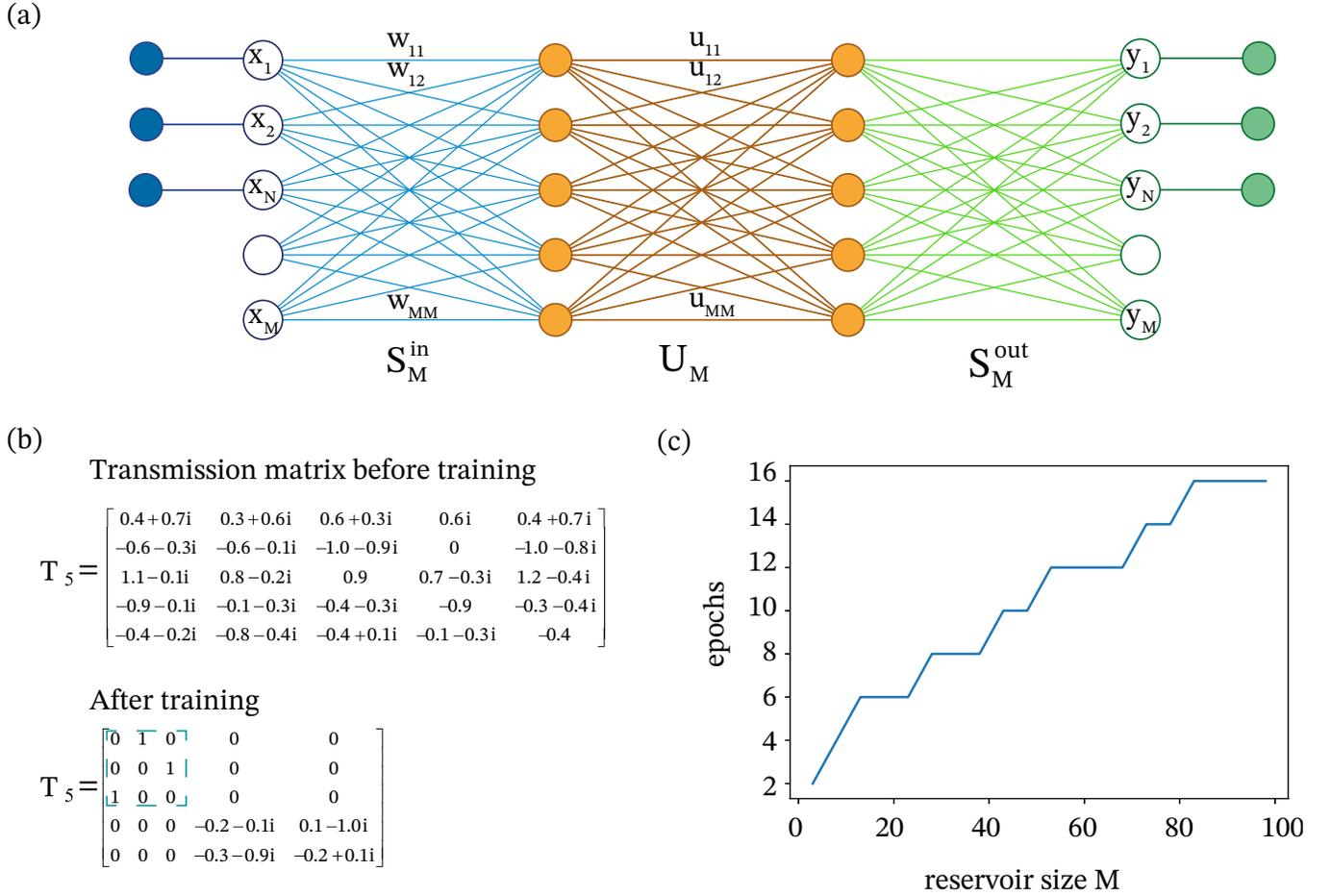} 
 \caption{Example of inference training of a random system ($M=5$) 
   to act as $X_3$ gate.
   (a) Neural network model (in our example $S_M^{\text{out}}$ is not used); (b) numerical examples for the trasmission
   matrix $T_M=U_M\cdot S_M^{\text{in}}$ before and after training; (c) scaling properties in terms of training epochs. Parameters: $n_{train}=100$,
   $n_{valid}=50$, $e_{valid}=10^{-3}$, $n_{epoch}=6$}.
  \label{fig:inference1}
\end{figure*}

Figure~\ref{fig:inference1} shows the training of a single qutrit gate $X_3$ in \eqref{eq:1}.
Similar results are obtained with other single qudit gates as $X^2$ and $Z$ and for higher dimensions. Training typically needs tens of iterations and scales well with the number of dimensions. Figure~\ref{fig:inference1} shows an example with $N=3$ and $M=5$. Figure~\ref{fig:inference1}c shows
that the number of training epochs $n_{epochs}$ scales linearly with the embedding space dimension $M$.
\section{Spatial light modulator implementation}
In the general case, one needs a unitary gate to train the complex medium, and a modulator to test different
inputs signals. In practical and simplified implementations, the training gate and the input modulator can be made with a single device. It is possible to realize the ML design with a single spatial light modulator (SLM), as sketched in the inset of Fig.~\ref{fig:SLMscaling}a. Ref.~\cite{Huisman:15} already gave a recipe for implementing a unitary in a lossy way with a single SLM and a complex medium. However, here we follow the more recent but also lossy technique introduced in Ref.~\cite{Matthes2018}. We consider an input plane wave represented by a constant vector ${\bf e}_N={1,1,...,1}$ with dimension $N$,  where $N$ is the number of pixels in the amplitude and phase SLM.

Assuming that we want to design a gate with input ${\bf x}$ and output ${\bf y}$, we generate the input ${\bf  x}$ by an operator $\mathrm{Diag}({\bf x})$, which has the first $N$ elements of ${\bf x}$ on the diagonal. 

Assuming that the ML algorithm has produced an operator $S_M$,
the actual operator to be implemented on the SLM is 
$\tilde S_M=S_M\cdot \mathrm{Diag}({\bf x})$. Note that $\tilde S_M$ encodes the input and hence changes for different inputs~\cite{Matthes2018}.

In other words, with a single SLM, after optimization for a given output  ${\bf y}$, the training realizes $\tilde S_M$ for a fixed plane wave input ${\bf e}_M={1,1,...,1,0,...0}$ with $N$ ones and $M-N$ zeros. 

\subsection{Phase-only modulators}
\begin{figure*}
  \centering
  \includegraphics[width=2.0\columnwidth]{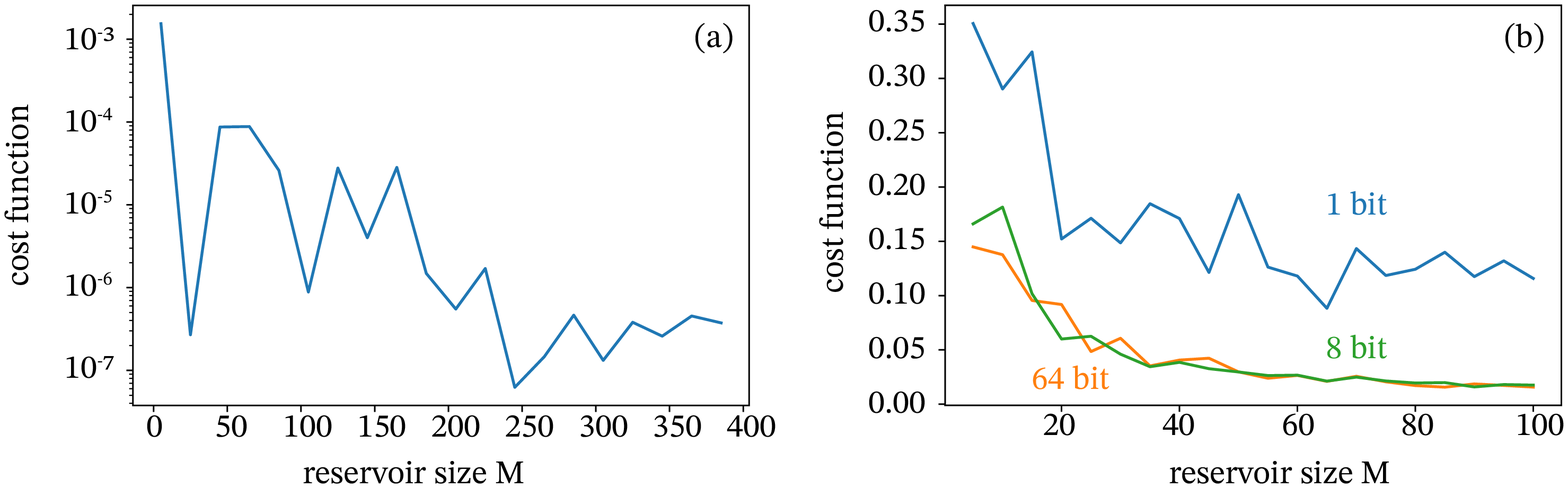}
  \caption{
    (a) Error after $1000$ training epochs versus the size of the reservoir $M$; the inset
    shows a sketch of the experimental implementation with a single spatial light modulator (SLM).
    (b) Error versus reservoir size $M$ after $1000$ epochs with a single amplitude modulator (with sign) with
    quantized levels (different bit numbers are indicated). \label{fig:SLMscaling}}
\end{figure*}

A pure phase modulator is implemented by the elements writing the model matrix for $\tilde S_M$ as $\cos(\phi_{ij})+\imath \sin(\phi_{ij})$,
with $\phi_{ij}$ the phase of the $i,j$ segment of the SLM.
In Fig.~\ref{fig:SLMscaling}a, we show the performance of the training process, focusing on a single qutrit $X-$gate ($N=3$), and varying the size of the reservoir $M$. If the reservoir is about one order of magnitude larger than the dimension of the gate, the algorithm converges in less than $1000$ epochs, and the error decreases with $M$.

\subsection{Sign modulators and quantized amplitude}
A pure amplitude modulator is modeled by a real matrix  $\tilde S_M$.
A combination of an amplitude modulator, such as a digital micromirror device (DMD), along with spatial filtering, enables one to
realize positive and negative values for $\tilde S_M$ \cite{Matthes2018}.
The elements of the real $\tilde S_M$ are trained to provide the target output with the fixed plane wave at the input.
Using typical functions in application program interfaces, such as {\it tensorflow.clip\textunderscore by\textunderscore value}, one can clip the values of the amplitude modulation (we use the range $[-1.0,1.0]$).
In contrast with the phase-modulator case, the performance in the amplitude modulation case is reduced.
Our numerical experiments show that convergence (corresponding to a cost-function smaller than $10^{-4}$) is not reached. On the contrary, the error reaches a stable minimal value after about $1000$ epochs.
The minimal error decreases with the size of the reservoir 
(Fig.~\ref{fig:SLMscaling}b). 
In Fig.~\ref{fig:SLMscaling}b, we also account for the fact that
modulator devices have limited resolution, and accessible modulation levels are quantized with a given number of bits.
We can implement the level quantization in TensorFlow by using 
{\it tensorflow.quantize\textunderscore and\textunderscore dequantize}, after each iteration. In Fig.~\ref{fig:SLMscaling}, we show results for phase-only modulation for the $1$-bit case, corresponding to modulation levels ${-1,0,1}$, as well as for the $8$ and $64$-bit cases.
\section{Conclusions}
We have investigated the use of machine learning paradigms for designing linear multi-level quantum
gates by using a complex transmitting multi-modal system.
The developed algorithms are versatile and scalable when the unitary operator for the random system is either known or unknown.
We show that generalized single-qudit gates can be designed.
The overall methodology is easily implemented by the TensorFlow application program interface and can be directly adapted to experimentally retrieved data. The method can be generalized to more complex information protocols, and embedded in real-world multi-modal systems.

{\it Acknowledgments --}
We acknowledge support from the Sapienza Ateneo,
PRIN2015 NEMO project (2015KEZNYM), the H2020 QuantERA project
QUOMPLEX (grant number 731473), and the PRIN 2017 PELM project
(20177PSCKT), the H2020 grant number 820392.

\clearpage
%


\end{document}